\documentclass[aps,prd,twocolumn]{revtex4}

\usepackage{amsmath}
\usepackage{graphicx}

\begin{document}

\title{Brane cosmological evolution with a general bulk matter configuration}
\author{Pantelis S. Apostolopoulos\footnote{Email address: papost@phys.uoa.gr} and Nikolaos Tetradis\footnote{Email address: ntetrad@cc.uoa.gr}}
\date{\today}

\begin{abstract}
Using a fully covariant treatment for the description of the bulk geometry,
we study the brane cosmological evolution in the presence of a smooth bulk
matter distribution. We focus on the case of a Friedmann-Robertson-Walker
(FRW) brane, invariantly characterized by the existence of a six-dimensional
group of isometries acting on 3D spacelike orbits. With a FRW brane, the
bulk geometry can be regarded as the 5D generalization of the \emph{%
inhomogeneous orthogonal family of Locally Rotationally Symmetric (LRS) }
spacetimes. We show that, for \emph{any bulk matter configuration}, the
expansion rate on the brane depends only on the covariantly defined \emph{%
comoving mass} $\mathcal{M}$ of the bulk fluid within a radius equal to the
average length scale of the 3D spacelike hypersurfaces of constant
curvature. This unique contribution incorporates the effects of the 5D Weyl
tensor and the projected tensor related to the bulk matter, and gives a
transparent physical picture that includes an effective conservation
equation between the brane and the bulk matter.
\end{abstract}

\address{University of Athens, Department of Physics, Nuclear and
Particle Physics Section, Panepistimiopolis, Zografos 15771, Athens, Greece.}
%\draft
%\preprint{HEP/123-qed}
\maketitle
\section{Introduction}

\setcounter{equation}{0}

The possibility that our Universe is identified with a 4D hypersurface
(3-brane) embedded in a higher-dimensional bulk space allows for interesting
novel features in the cosmological evolution as perceived by a brane
observer. Consistency with the standard cosmological expansion can be
achieved if the extra dimensions are ``warped'' in a way that low-energy
gravitons are localized on the brane \cite{rs}. This requires a negative
cosmological constant in the bulk. The presence of matter localized on the
brane induces a cosmological expansion. The 3-brane must also have a tension
that balances the effect of the negative cosmological constant on the
expansion. For low energy densities, the effective Friedmann equation on the
brane has the standard form \cite{binetruy,csaki,kraus,maeda}. Several
features and variations of this scenario have been consider. (For recent
reviews see \cite{brax,Maartens1}.) We are interested in the effects arising
from the presence of additional matter in the bulk, such as gravitational
radiation emitted or absorbed by the brane \cite{vaidya,tet2}, or bulk
fields \cite{dilaton}. The possible energy exchange between the brane and
the bulk can induce important modifications of the cosmological evolution 
\cite{vandebruck,tet1,tetap}.

In this work we would like to address the problem of the effect of the bulk
matter on the brane cosmological evolution in a general way. For this reason
we employ the covariant formalism \cite{Maartens1,maeda} that permits us to
describe the evolution without reference to a particular local system of
coordinates. We assume that the spatial part of the brane metric is
maximally symmetric, in agreement with the homogeneity and isotropy of the
observed Universe at large scales. We refer to it as a
Friedmann-Robertson-Walker (FRW) brane. As a consequence, the bulk geometry
can be seen as the 5D generalization of the \emph{inhomogeneous orthogonal
Locally Rotationally Symmetric (LRS)} spacetimes \cite{elst-ellis}. Our main
result is that the symmetry of the FRW brane constraints the form of the
bulk contribution to the brane evolution. The effective Friedmann equation
contains only one term originating in the bulk matter. The physical quantity
affecting the brane evolution is the covariantly defined \emph{comoving mass}
of the bulk fluid within a radius equal to the average length scale of the
3D spacelike hypersurface (essentially the value of the scale factor).

In the following section we summarize the known results for the bulk-brane
dynamics within the covariant formulation. In section III we concentrate on
the case of a FRW brane. We invariantly characterize the bulk geometry in
terms of the unique preferred spacelike direction that represents the local
(axis) of symmetry. This naturally emerges from the assumption of the
rotational symmetry of the 5D bulk. In section IV we express covariantly the
electric part of the Weyl tensor in terms of the irreducible parts of a
general bulk energy-momentum tensor with respect to a generic bulk observer $%
u^A$. The comoving mass appears as a first integral of the constraint
equations satisfied by the shear tensor and the electric part of the 5D Weyl
tensor. As two concrete and illustrative examples, we apply the above
considerations to the Gauss-normal coordinate system in which the brane is
located at a fixed value of the fourth spatial coordinate (\emph{brane
comoving system}), and to the coordinate system adapted to the bulk
observers $u^A$ (\emph{bulk comoving system}) in which the brane is moving.
This approach shows the equivalence of these points of view for \emph{any}
bulk matter configuration. In section V we use these results in order to
derive the effective Friedmann and Raychaudhuri equations governing the
cosmological evolution on the FRW brane. A combination of these equations
has the form of a conservation equation involving the brane energy density
and an effective density originating in the bulk matter. Our conclusions are
given in section VI.

Throughout this work we use the following index conventions: bulk 5D indices
are denoted by capital latin letters $A,B,...=0,1,2,3,4$ and greek letters
denote spacetime indices $\alpha ,\beta ,...=0,1,2,3$.
%and lower latin letters correspond to 3D coordinates $i,j,...=1,2,3$.

\section{Brane dynamics with matter in the bulk}

\setcounter{equation}{0} We are interested in cases with a matter component
in the bulk in addition to the negative cosmological constant. We consider
an action of the form 
\begin{eqnarray}
S &=&\int d^5x\sqrt{-g}\left( \Lambda +M^3R+\mathcal{L}_{\text{\textsc{bulk}}%
}^{mat}\right) +  \nonumber \\
&&  \nonumber \\
&&+\int d^4x\sqrt{-g}\,\left( -V+\mathcal{L}_{\text{\textsc{brane}}%
}^{mat}\right) ,  \label{action1}
\end{eqnarray}
where $R$ is the curvature scalar of the five-dimensional bulk metric $%
g_{AB} $, $-\Lambda $ the bulk cosmological constant ($\Lambda >0$), $V$ the
brane tension, and $g_{\alpha \beta }$ the induced metric on the brane. The
brane is assumed to be normal to the unit spacelike vector field $n^A$ that
is tangential to the extra spatial dimension.

The Einstein Equations (EE) in the bulk take the usual form 
\begin{equation}
G_{~B}^A=\frac 1{2M^3}\left( T_{~B}^A+\Lambda \delta _{~B}^A\right),
\label{einstein}
\end{equation}
where $T_{~B}^A$ denotes the total energy-momentum tensor, i.e. 
\begin{equation}
T_{AB}=T_{AB}^{\text{\textsc{bulk}}}+\delta \left( n\right) \left( T_{AB}^{%
\text{\textsc{brane}}}+Vg_{AB}\right).  \label{energy-momentum1}
\end{equation}
The modified 4D EE are derived by assuming a $Z_2$ symmetry of the bulk
around the brane and employing Israel's junction conditions \cite
{israel,Maartens1} 
\begin{equation}
G_{\alpha \beta }=-3\lambda g_{\alpha \beta }+\frac V{24M^6}T_{\alpha \beta
}+\frac 1{4M^6}\mathcal{S}_{\alpha \beta }-\mathcal{E}_{\alpha \beta }+\frac
1{3M^3}\mathcal{F}_{\alpha \beta },  \label{branefieldequations}
\end{equation}
where 
\begin{equation}
\mathcal{S}_{\alpha \beta }=\frac 12TT_{\alpha \beta }-\frac 14T_{\alpha
\gamma }T_{~\beta }^\gamma +\frac{3T_{\gamma \delta }T^{\gamma \delta }-T^2}{%
24}g_{\alpha \beta }  \label{def1}
\end{equation}
\begin{equation}
\mathcal{F}_{\alpha \beta }=T_{AB}^{\text{\textsc{bulk}}}g_{\hspace{0.15cm}%
\alpha }^Ag_{\hspace{0.15cm}\beta }^B+\left( T_{AB}^{\text{\textsc{bulk}}%
}n^An^B-\frac{T^{\text{\textsc{bulk}}}}4\right) g_{\alpha \beta }
\label{def2}
\end{equation}
\begin{equation}
\mathcal{E}_{\alpha \beta }=\mathcal{E}_{AB}g_{\hspace{0.15cm}\alpha }^Ag_{%
\hspace{0.15cm}\beta }^B=C_{ACBD}n^Cn^Dg_{\hspace{0.15cm}\alpha }^Ag_{%
\hspace{0.15cm}\beta }^B  \label{Weyl1}
\end{equation}
and $\lambda =(V^2/12M^3-\Lambda )/12M^3$ is the effective cosmological
constant.

From the EE (\ref{branefieldequations}) we observe that, apart from the
terms quadradic to the brane energy-momentum tensor, there exist two
additional terms corresponding to: a) the projection of the 5D Weyl tensor $%
\mathcal{E}_{\alpha \beta }$ and b) the projected (normal to $n^A$ ) 
tensor $\mathcal{F}_{\alpha \beta }$ that contains the bulk matter contribution. Since both
tensors are 5D objects we conclude that \emph{both induce bulk effects on
the brane}. In the case of an empty bulk, the 5D contributions on the brane
are coming from the non-local effects of the free gravitational field
incorporated in $\mathcal{E}_{\alpha \beta }$ (5D bulk gravitons). As we
shall see in the subsequent sections, when the bulk is non-empty, the tensor $\mathcal{E}_{\alpha \beta }$ 
is expressed in terms of the bulk matter configuration. Therefore it is clear that, in order to analyze the
cosmological evolution on the brane, we must take into account the
contribution of the bulk energy-momentum tensor.

Formally \emph{any type }of bulk matter can be covariantly described by an
energy-momentum tensor decomposed with respect to the cosmological (brane)
observers $\tilde{u}_\alpha \equiv g_{\hspace{0.15cm}\alpha }^A$ $\tilde{u}%
_A $ as\ 
\begin{equation}
T_{AB}^{\text{\textsc{bulk}}}=\bar{\rho}\tilde{u}_A\tilde{u}_B+\bar{p}\tilde{%
h}_{AB}+2\bar{q}_{(A}\tilde{u}_{B)}+\bar{\pi}_{AB}.  \label{energy-decomp1}
\end{equation}
The energy density $\bar{\rho}$, isotropic pressure $\bar{p}$, energy flux
vector $\bar{q}_A$ and anisotropic pressure tensor $\bar{\pi}_{AB}$,
measured by the brane observers, are \emph{\ covariantly defined} as 
\begin{eqnarray}
\bar{\rho} &=&T_{AB}\tilde{u}^A\tilde{u}^B,\text{ }\bar{p}=\frac 14T_{AB}%
\tilde{h}^{AB},\text{ }\bar{q}_A=-\tilde{h}_A^CT_{CD}\tilde{u}^D,  \nonumber
\\
&&  \nonumber \\
\bar{\pi}_{AB} &=&\tilde{h}_A^C\tilde{h}_B^DT_{CD}-\frac 14(\tilde{h}%
^{CD}T_{CD})\tilde{h}_{AB},  \label{dynam-quantities}
\end{eqnarray}
where $\tilde{h}_{AB}=g_{AB}+\tilde{u}_A\tilde{u}_B$ is the projection
tensor perpendicularly to $\tilde{u}^A$.

With these identifications the conservation equation on the brane, coming
from the twice contracted Bianchi identities, implies \cite{Maartens1} 
\begin{equation}
T_{\alpha \beta }^{\hspace{0.4cm};\beta }=-2\left[ \left( \bar{q}%
_Cn^C\right) \tilde{u}_\alpha +\bar{\pi}_{AB}n^Bg_{\hspace{0.15cm}\alpha
}^A\right].  \label{Bianchi-brane}
\end{equation}
This shows that in general the brane matter is not conserved, but there is 
\emph{energy exchange} (outflow or inflow) between the brane and bulk, that
depends on the character of the vector field $\left( \bar{q}_Cn^C\right) 
\tilde{u}_\alpha +\bar{\pi}_{AB}n^Bg_{\hspace{0.15cm}\alpha }^A$ involving
the energy flux vector $\bar{q}_A$ and the bulk anisotropic stress vector $\bar{%
\pi}_{AB}n^B$.

\section{An invariant description of the bulk geometry with a FRW brane}

For a FRW brane the 3D hypersurfaces $\mathcal{D}$ normal to the \emph{prolongated}
cosmological observers $\tilde{u}_A$ are maximally symmetric. In geometrical
terms this means that the full 5D space admits a six-dimensional multiply
transitive group of isometries acting on 3D spacelike orbits. This in turn
implies the existence of a one-parameter continuous isotropy subgroup.
Consequently, the bulk geometry with a FRW brane can be seen as a 5D
generalization of the \emph{inhomogeneous orthogonal family of Locally
Rotationally Symmetric} (\emph{LRS class II}) spacetimes \cite{elst-ellis}.

Let us examine the kinematics and dynamics of the bulk from the point of
view of a bulk observer $u^A$ ($u^Au_A=-1$ and $u_A\neq \tilde{u}_A$). The
deformation of the timelike congruence $u^A$ is covariantly described by the
irreducible 1+4 threading of its first derivatives in the standard way \cite
{ellis} 
\begin{equation}
u_{A;B}=\Sigma _{AB}+\frac \Theta 4h_{AB}+\Omega _{BA}-\dot{u}_Au_B,
\label{bulk-kin-quant}
\end{equation}
where $\Sigma _{AB}=\left (h^K_Ah^L_B-\frac{1}{4}h^{KL}h_{AB}\right)u_{(K;L)}$, 
$\Theta =u_{;A}^A$, $\Omega _{BA}=u_{[A;B]}+\dot{u}_{[A}u_{B]}$ and 
$\dot{u}^A=u_{;B}^Au^B$ are the rate of shear tensor, the
rate of expansion scalar, the vorticity tensor and the acceleration of the
observers $u^A$, respectively, and $h_{AB}=g_{AB}+u_Au_B$ is the projection
operator perpendicularly to $u^A$. We assume that the timelike congruence $%
u^A$ is irrotational, i.e. $\Omega _{AB}=0$. This is always possible within
the geometry we are considering.

Regarding the dynamics, the matter content of the bulk is described by the
energy-momentum $T_{AB}^{\text{\textsc{bulk}}}$, which with respect to the
observers $u^A$ can be written as 
\begin{equation}
T_{AB}^{\text{\textsc{bulk}}}=\rho u_Au_B+ph_{AB}+2q_{(A}u_{B)}+\pi _{AB}.
\label{energy-decomp2}
\end{equation}
The dynamical quantities measured by the bulk observers are defined in a
similar manner as in (\ref{dynam-quantities}).

Because of the LRS bulk geometry, there exists a \emph{preferred spacelike
direction} $e^A$ representing the (local) axis of symmetry with respect to
which all the geometrical, kinematical and dynamical quantities are
invariant. It is natural to select $e^A$ to be the normal to the timelike
congruence of the bulk observer $u^A$, i.e. $u^A e_A=0$. As a result all the
spacelike vector or tensor fields perpendicular to $u^A$ can be written in
terms of $e^A$.

In complete analogy with \cite{elst-ellis} we introduce the \emph{unique}
spacelike and trace-free tensor 
\begin{equation}
e_{AB}=e_{(AB)}=\frac 13\left( 4e_Ae_B-h_{AB}\right) =h_{AB}-\frac 43\Pi
_{AB}  \label{trace-free}
\end{equation}
where the projection operator $\Pi _{AB}$ is defined as 
\begin{equation}
\Pi _{AB}\equiv g_{AB}+u_Au_B-e_Ae_B=h_{AB}-e_Ae_B  \label{1+1+3operator}
\end{equation}
and has the usual projection properties

\begin{equation}
\Pi _A^A=3,\qquad \Pi _C^A\Pi _B^C=\Pi _B^A,\qquad \Pi _B^Ae^B=\Pi _B^Au^B=0,
\label{properties}
\end{equation}
i.e. it projects normally both to $u^A$ and $e^A$. Essentially $\Pi _{AB}\,$%
is used for the 1+1+3 decomposition of the tangent space. In the particular
case of the FRW brane it corresponds to the induced metric of the 3D
hypersurfaces of constant curvature.

Using the above definitions it will be convenient to express every covariantly defined spacelike and
traceless tensor $\Lambda _{AB}$ as 
\begin{equation}
\Lambda _{AB}=\sqrt{\frac 32}\Lambda e_{AB}  \label{trace-free-tensors}
\end{equation}
where $\Lambda ^2={\Lambda _{AB}\Lambda ^{AB}}/2$ is the norm of the
trace-free tensor $\Lambda _{AB}$.\newline
\newline

\section{The Weyl radiative energy}

Our aim in this section is to express the scalar 
\begin{equation}
\mathcal{E}=C_{ACBD}\tilde{u}^An^C\tilde{u}^Bn^D  \label{electric-scalar}
\end{equation}
in terms of the bulk matter. It will prove easier to compute the scalar $%
E=C_{ACBD}u^Ae^Cu^Be^D$ with respect to the bulk observers $u^A$ and the
preferred spatial direction $e^A$ ($e^Au_A=0$). It can be shown easily, by
performing a 1+4 decomposition of $\tilde{u}^A$ and $n^A$ w.r.t. $u^A$ and $%
e^A$, that $\mathcal{E}=E$.

Using the Ricci identities for $u^A$, the normal projection of its trace and
the energy-momentum decomposition (\ref{energy-decomp2}) gives the
generalized \emph{constraint equation for the shear tensor} of the
(irrotational) bulk observers 
\begin{equation}
D_K\Sigma _{\hspace{0.2cm}A}^K=\frac 34D_A\Theta -\frac 1{2M^3}q_A.
\label{constraint1}
\end{equation}
We have employed here the fully projected (perpendicular to $u^A$) covariant
derivative $D_A$ which, for any tensor $P_{\hspace{0.35cm}\hspace{0.35cm}%
IJ...}^{AB...}$, is defined as 
\begin{equation}
D_LP_{\hspace{0.35cm}\hspace{0.35cm}IJ...}^{AB...}\equiv
h_R^Ah_S^B...h_I^Th_J^X...h_L^K\left( P_{\hspace{0.35cm}\hspace{0.35cm}%
TX...}^{RS...}\right) _{;K}  \label{projected-derivative}
\end{equation}
where a semicolon ``;" denotes the usual covariant derivative associated
with the bulk metric $g_{AB}$.

The trace of Bianchi identities $R_{AB[CD;E]}=0$ in a 5D space gives 
\begin{equation}
C_{\hspace{0.2cm}ABC;K}^K=\frac 43\left( R_{A[C;B]}-\frac
18g_{A[C}R_{;B]}\right) .  \label{trace-Bianchi-ident}
\end{equation}
We note that, in contrast to the 4D geometries, equation (\ref
{trace-Bianchi-ident}) \emph{is not equivalent} to the full set of Bianchi
identities. However, for the purposes of the present work it will be
sufficient. In fact, the 1+4 decomposition of (\ref{trace-Bianchi-ident})
gives propagation (parallel to $u^A$) and constraint (normal to $u^A$)
equations for the electric part $E_{AB}=C_{ACBD}u^Cu^D$ of the Weyl tensor.
From these we shall need only the latter. After a straightforward
calculation we obtain the generalized \emph{constraint equation for the
electric part} of the 5D Weyl tensor 
\begin{widetext}
\begin{equation}
D_K\left( E_{\hspace{0.2cm}A}^K+\frac 1{2M^3}\frac 23\pi _{\hspace{0.2cm}%
A}^K\right) =\frac 1{2M^3}\frac 12D_A\rho +\frac 1{2M^3}\frac 23\Sigma _{%
\hspace{0.2cm}A}^Kq_K-\frac 1{2M^3}\frac 12q_A\Theta -h_A^BC_{\hspace{0.2cm}%
LCB}^K\Sigma _{\hspace{0.2cm}K}^Cu^L,  \label{constraint-electric1}
\end{equation}
\end{widetext}
where we have incorporated the EE (\ref{einstein}) \emph{away from the brane}.

Denoting with a prime the spatial derivative with respect to the preferred
direction $e^A$, and using the fact that $\Sigma _{\hspace{0.2cm}B}^A$ and $%
\pi _{\hspace{0.2cm}B}^A$ are spacelike traceless tensors, equations (\ref
{trace-free-tensors}), (\ref{constraint1}) and (\ref{constraint-electric1})
give 
\begin{equation}
\frac 1{\ell ^4}\left\{ \left[ E+\frac 1{2M^3}\frac 23\left( \pi
_{AB}e^Ae^B\right) \right] \ell ^4\right\} ^{\prime }=\frac 1{2M^3}\frac
12\rho ^{\prime }.  \label{constraint-electric2}
\end{equation}
We have introduced the average length scale function $\ell $ of the
spacelike congruence $e^A$ defined as $D_Ae^A\equiv 3\ell ^{\prime }/\ell $. Essentially the quantity $D_Ae^A$ corresponds to the overall
expansion of the spacelike congruence as measured in the rest space of the
observers $u^A$. Therefore, $\ell$ represents the \emph{radius of the 3D spatial slices} 
$\mathcal{D}$.    

Equation (\ref{1+1+3operator}) implies 
\begin{equation}
\pi _{AB}e^Ae^B=\left( T_{AB}^{\text{\textsc{bulk}}}-ph_{AB}\right)
e^Ae^B=T^{\text{\textsc{bulk}}}+\rho -3p_{\perp }-p.  \label{auxiliary1}
\end{equation}
The pressure $3p_{\perp }\equiv \Pi ^{AB}T_{AB}^{\text{\textsc{bulk}}}$ is
perpendicular to both $u^A$,$e^A$ and corresponds to the \emph{isotropic
pressure} measured by the bulk observers. 

Integrating (\ref{constraint-electric2}) and using equation (\ref{auxiliary1}%
) we finally get 
\begin{equation}
\mathcal{E}=E=-\frac 1{2M^3}\left[ \frac 12\left( T^{\text{\textsc{bulk}}%
}-4p_{\perp }\right) +\frac{\mathcal{M}}{\pi ^2\ell ^4}\right] ,
\label{electric1}
\end{equation}
where 
\begin{equation}
\mathcal{M}=\int_0^\ell 2\pi ^2\rho \ell ^3d\ell +\mathcal{M}_0
\label{mass-function}
\end{equation}
can be interpreted as the generalized \emph{comoving mass} of the bulk fluid
within radius $\ell $.

It would be useful to give some familiar examples by choosing specific
(local) coordinate systems, in order to express equation (\ref{electric1})
in analytic form. In particular, it is possible to discuss the brane
evolution either in the Gauss-normal coordinate system in which the brane is
located at a fixed value of the fourth spatial coordinate (\emph{brane
comoving system}), or in the coordinate system adapted to the bulk observers 
$u^A$ (\emph{bulk comoving system}) in which the brane is moving \cite
{binetruy, csaki, kraus}.

In the latter case local coordinates can be found such that $%
u^A=n^{-1}(t,r)\delta _{~~t}^A$ and the bulk metric is written in \emph{%
Schwarzschild coordinates} as 
\begin{equation}
ds^2=-n^2(t,r)dt^2+b^2(t,r)dr^2+r^2d\Omega _k^2  \label{metric1}
\end{equation}
where $d\Omega _k^2$ is the metric of the 3D hypersurfaces $\mathcal{D}$ 
of constant curvature that is parametrized by $k=-1,0,1$.   
Obviously the preferred spatial axis of symmetry is $\sim \partial _r$, and
therefore $\ell \equiv r$. This implies that the integrated matter
distribution between $r=0$ and some arbitrary radius $r$ is $\mathcal{M}%
=\int_0^r2\pi ^2\rho r^3dr+\mathcal{M}_0$. The integration constant $%
\mathcal{M}_0$ is the contribution from the mass of a black hole located at $%
r=0$.

On the other hand, using Gauss-normal coordinates for the bulk with the
brane located at a fixed value (e.g. $\eta =0$) of the fourth spatial
coordinate, the metric is 
\begin{eqnarray}
ds^2 &=&\gamma _{\alpha \beta }dx^\alpha dx^\beta +d\eta ^2=  \nonumber \\
&&-{\tilde{m}}^2(\tilde{\tau},\eta )d\tilde{\tau}^2+R^2(\tilde{\tau},\eta
)d\Omega _k^2+d\eta ^2.  \label{Gauss-normal}
\end{eqnarray}
In this case $\ell = R(\tilde{\tau},\eta )$ represents the average length 
scale for distances between any pair of brane observers. It follows 
that, at the location of the brane $\eta =0$, the quantity $\ell$ corresponds to \emph{the scale factor} of the 
FRW brane. The generalized comoving mass is $\mathcal{M}=\int_0^R2\pi ^2\rho R^3dR+\mathcal{M}_0
=\int_0^R2\pi ^2T_{AB}^{\text{\textsc{bulk}}}u^Au^BR^3dR+\mathcal{M}_0$.

We conclude this section by pointing out that the equivalence between the
two points of view has been shown for an empty bulk ($T_{AB}^{\text{\textsc{%
bulk}}}=0$). In this case the 5D metric is static and reduces to the
well-known 5D AdS-Schwarzschild with $\mathcal{M}=\mathcal{M}_0$
representing the black hole mass \cite{kraus,birm,gregory}. The present
approach shows that this equivalence holds in general and \emph{%
irrespectively of the specific assumptions for the bulk matter distribution}.

\section{The induced modifications of the brane evolution}

The invariant characterization of the FRW brane is the vanishing of the
vorticity, shear and acceleration of the timelike congruence $\tilde{u}%
_\alpha $. Assuming a perfect fluid matter configuration on the brane, the
energy-momentum tensor is written in terms of the energy density $\tilde{\rho%
}$ and the isotropic pressure $\tilde{p}$ as 
\begin{equation}
T_{\alpha \beta }=\tilde{\rho}\tilde{u}_\alpha \tilde{u}_\beta +\tilde{p}%
\tilde{h}_{\alpha \beta }.  \label{energy_momentum_brane}
\end{equation}
The conservation equation (\ref{Bianchi-brane}) can be split along and
normally to $\tilde{u}^\alpha $. In this way we obtain the propagation
equation for the energy density $\tilde{\rho}$ 
\begin{equation}
\stackrel{\cdot }{\tilde{\rho}}+3H\left( \tilde{\rho}+\tilde{p}\right) =2%
\bar{q}_Cn^C .  \label{propagation-density}
\end{equation}
This equation shows that the rate of energy transfer between the bulk and
the brane is controlled by the energy flux vector field $\bar{q}_A$ that
appears in the decomposition of the bulk energy-momentum tensor according to
(\ref{energy-decomp2}).

It should be noted that the normal part of (\ref{Bianchi-brane}) gives the
spatial gradient of $\tilde{p}$

\begin{equation}
h_\alpha ^\beta \tilde{p}_{;\beta }=-2\bar{\pi}_{AB}n^Bg_{\hspace{0.15cm}%
\alpha }^A.  \label{pressure-gradient}
\end{equation}
However, taking into account the relation (\ref{trace-free-tensors}),
equation (\ref{pressure-gradient}) implies the standard spatial homogeneity $%
h_\alpha ^\beta \tilde{p}_{;\beta }=0$ of the isotropic pressure $\tilde{p}$.

The brane evolution can be studied by determining the generalized Friedmann
and Raychaudhuri equations \emph{on the brane} in the presence of bulk
matter. These follow from the Gauss-Codazzi equations and the timelike part
of the trace of the Ricci identities applied to the (irrotational, geodesic
and shear-free) timelike congruence $\tilde{u}_\alpha $. They have the form 
\begin{equation}
H^2=\frac 13R_{\alpha \beta }\tilde{u}^\alpha \tilde{u}^\beta -\frac{^3R}%
6+\frac R6  \label{Friedmann1}
\end{equation}
\begin{equation}
\dot{H}=-H^2-\frac 13R_{\alpha \beta }\tilde{u}^\alpha \tilde{u}^\beta,
\label{Raychaudhuri1}
\end{equation}
where $R_{\alpha \beta }=G_{\alpha \beta }-\frac G2g_{\alpha \beta }$ is the
modified Ricci tensor of the brane, $3H=\tilde{u}_{;\alpha }^\alpha $ the
Hubble parameter, $^3R$ the scalar curvature of the 3D hypersurfaces $%
\mathcal{D}$, and a dot denotes differentiation with respect to $\tilde{u}%
^\alpha$, i.e. $\dot{H}=H_{;\alpha }\tilde{u}^\alpha $.

Using equation (\ref{branefieldequations}) and the decomposition (\ref
{energy-decomp1}), equations (\ref{Friedmann1}) and (\ref{Raychaudhuri1})
become:

\begin{eqnarray}
H^2 &=&\lambda -\frac{^3R}6+\frac{\left( \tilde{\rho}^2+2V\tilde{\rho}%
\right) }{144M^6}-  \nonumber \\
&-&\left\{ \frac{\mathcal{E}}3-\frac 1{12M^3}\left[ \bar{\rho}-\frac
43\left( \bar{p}+T_{AB}^{\text{\textsc{bulk}}}n^An^B\right) \right] \right\}
\label{Friedmann}
\end{eqnarray}
\begin{eqnarray}
\dot{H} &=&-H^2+\lambda -\frac{V\left( \tilde{\rho}+3\tilde{p}\right) }{%
144M^6}-\frac{\left( 2\tilde{\rho}^2+3\tilde{\rho}\tilde{p}\right) }{144M^6}-
\nonumber \\
&-&\left\{ \frac 1{12M^3}\left[ \bar{\rho}+\frac 43\left( \bar{p}+\frac
12T_{AB}^{\text{\textsc{bulk}}}n^An^B\right) \right] -\frac{\mathcal{E}}%
3\right\} .  \label{Raychaudhuri}
\end{eqnarray}
with $\mathcal{E}=\mathcal{E}_{\alpha \beta }\tilde{u}^\alpha \tilde{u}%
^\beta =C_{ACBD}\tilde{u}^An^C\tilde{u}^Bn^D$.

Employing the corresponding projection tensor $\bar{\Pi}%
_{AB}=g_{AB}+\tilde{u}_A\tilde{u}_B-n_An_B$ (projecting normally to $n^A$
and $\tilde{u}^A$), we decompose the isotropic pressure $\bar{p}$ into
parallel and perpendicular parts according to 
\begin{equation}
\bar{p}_{\parallel }=T_{AB}^{\text{\textsc{bulk}}}n^An^B,\hspace{0.4cm}%
p_{\perp }=\frac 13T_{AB}^{\text{\textsc{bulk}}}\bar{\Pi}^{AB},\hspace{0.4cm}%
\bar{p}=\frac{\bar{p}_{\parallel }+3p_{\perp }}4  \label{pressure}
\end{equation}
\begin{equation}
\bar{\rho}=-T^{\text{\textsc{bulk}}}+\bar{p}_{\parallel }+3p_{\perp }.
\label{energy}
\end{equation}
We observe that $\bar{p}$ consists of two terms:\ a contribution $\bar{p}%
_{\parallel }$ along the direction of $n^A$, and the perpendicular part $%
p_{\perp }$ that corresponds to the isotropic bulk pressure parallel to the
3D hypersurfaces $\mathcal{D}$. We note that, in a general brane, there will
be an anisotropic bulk pressure described by the traceless part of $\bar{\Pi}%
_A^C\bar{\Pi}_B^DT_{CD}^{\text{\textsc{bulk}}}$. However, for a FRW brane,
because of the spatial homogeneity and isotropy of $\mathcal{D}$, the \emph{%
only pressure contribution coming from directions normal to both } $\tilde{u}%
^A$\emph{\ and }$n^A$\emph{\ is the isotropic part }$p_{\perp }$.

Inserting equations (\ref{pressure}) and (\ref{energy}) in (\ref{Friedmann})
and (\ref{Raychaudhuri}) we obtain: 
\begin{eqnarray}
H^2 &=&\lambda -\frac{^3R}6+\frac{\left( \tilde{\rho}^2+2V\tilde{\rho}%
\right) }{144M^6}-  \nonumber \\
&-&\left\{ \frac{\mathcal{E}}3-\frac 1{12M^3}\left[ 4p_{\perp }-T^{\text{%
\textsc{bulk}}}\right] \right\}  \label{Friedmann2}
\end{eqnarray}
\begin{eqnarray}
\dot{H} &=&-H^2+\lambda -\frac{V\left( \tilde{\rho}+3\tilde{p}\right) }{%
144M^6}-\frac{\left( 2\tilde{\rho}^2+3\tilde{\rho}\tilde{p}\right) }{144M^6}-
\nonumber \\
&-&\left\{ \frac 1{12M^3}\left[ 4p_{\perp }+2\bar{p}_{\parallel }-T^{\text{%
\textsc{bulk}}}\right] -\frac{\mathcal{E}}3\right\} .  \label{Raychaudhuri2}
\end{eqnarray}
Equations (\ref{Friedmann2}) and (\ref{Raychaudhuri2}) show how the brane
evolution is affected by the presence of matter in the bulk. For example, in
the case of a general bulk perfect fluid (w.r.t. to the bulk observers $u^A$%
) the normal pressure $p_{\perp }$ is \emph{equal} to the isotropic pressure 
$p$ measured by the bulk observers. Since $T^{\text{\textsc{bulk}}}=-\rho
+4p=-\rho +4p_{\perp }$, the last term in equation (\ref{Friedmann2})
becomes $\rho /12M^3$. The influence of the bulk fluid on the expansion
arises through the local bulk matter density $\rho$ and the scalar $\mathcal{%
E}$ that can be loosely interpreted as accounting for the effect of the
local gravitational field. In the case of a null radiation fluid (i.e. a
generalized AdS-Vaidya bulk) the trace of the bulk energy-momentum tensor
vanishes and the last term becomes $p_{\perp }/3M^3$.

Inserting equation (\ref{electric1}) in (\ref{Friedmann2}) and (\ref
{Raychaudhuri2}) we get:

\begin{equation}
H^2\equiv\frac {\dot \ell} {\ell}=\lambda -\frac{^3R}6+\frac{\left( \tilde{\rho}^2+2V\tilde{\rho}\right) }{%
144M^6}+\frac{\mathcal{M}}{6\pi ^2M^3\ell^4}  \label{Friedmann3}
\end{equation}

\begin{eqnarray}
\dot{H} &=&-H^2+\lambda -\frac{V\left( \tilde{\rho}+3\tilde{p}\right) }{%
144M^6}-\frac{\left( 2\tilde{\rho}^2+3\tilde{\rho}\tilde{p}\right) }{144M^6}-
\nonumber \\
&&  \nonumber \\
&&-\frac{\mathcal{M}}{6\pi ^2M^3\ell^4}-\frac 1{6M^3}\bar{p}_{\parallel }.
\label{Raychaudhuri3}
\end{eqnarray}
We conclude that the total bulk contribution to the Friedmann equation is
encompassed by the generalized \emph{comoving mass} $\mathcal{M}$ of the
bulk fluid, \emph{irrespectively of the dynamical interpetation with respect
to the bulk observers}.

Another intuitive reformulation of (\ref{Friedmann3}) and (\ref
{Raychaudhuri3}) is obtained by taking the derivative of the first equation
along $\tilde{u}^\alpha$ and combining the result with the second one. We
obtain the effective conservation equation 
\begin{equation}
\left[ \stackrel{\cdot }{\tilde{\rho}}+3H\left( \tilde{\rho}+\tilde{p}%
\right) \right] \left( 1+\frac{\tilde{\rho}}V\right) =-\left[ \dot{\rho}_{%
\text{eff}}+3H\left( \rho _{\text{eff}}+p_{\text{eff}}\right) \right],
\label{conservation-bulk-brane}
\end{equation}
where 
\begin{equation}
\rho _{\text{eff}}=\frac{12M^3}V\frac{\mathcal{M}}{\pi ^2\ell^4},\hspace{0.5cm}%
p_{\text{eff}}=\frac{\rho _{\text{eff}}}{3}+ \frac{8M^3}V\bar{p}_{\parallel }
\label{effective-matter-pressure}
\end{equation}
can be regarded as the \emph{effective energy density and pressure} arising
through the bulk matter. The effective Friedmann equation (\ref{Friedmann3})
can be written as 
\begin{equation}
H^2=\lambda -\frac{^3R}6+\frac{1}{6 M^2_{\mathrm{Pl}}} \left[\tilde{\rho}%
\left(1+ \frac{\tilde{\rho}}{2V}\right)+\rho _{\text{eff}} \right],
\label{Friedmann4}
\end{equation}
with $M^2_{\mathrm{Pl}}=12M^6/V$. For low brane energy density $\tilde{\rho}%
\ll V$, the terms $\sim \tilde{\rho}^2, \tilde{\rho}\tilde{p}$ become
negligible.

\section{Conclusions}

Our main results are summarized by equations (\ref{Friedmann3})--(\ref
{Friedmann4}). The first equation indicates that the effect of the bulk
matter on the Friedmann equation is incorporated in the generalized \emph{%
comoving mass} $\mathcal{M}$ of the bulk fluid, \emph{irrespectively of the
nature of this fluid}. This is reminiscent of the implications of Birkhoff's
theorem for the gravitational field generated by a matter distribution. Both
results are consequences of the assumed Local Rotational Symmetry of the
geometry that is inherited by the matter distribution (through the assumed 
symmetry inheritance by the bulk observers). 

The equivalent form (\ref{Friedmann4}) of the Friedmann equation indicates
that the bulk generates an effective matter component that contributes to
the brane cosmological evolution. For low brane energy densities two fluids
contribute linearly to the expansion: The brane matter with energy density $%
\tilde{\rho}$, and the effective (or ``mirage'') matter with energy density $%
\rho _{\text{eff}}$ given by the first of equations (\ref
{effective-matter-pressure}). A remarkable conclusion can be drawn from
equation (\ref{conservation-bulk-brane}) for low brane energy densities:
There is conservation of energy between the two components. The brane matter
scales as a perfect fluid with an equation of state $\tilde{p}=\tilde{p}(%
\tilde{\rho})$. The mirage component again scales as a perfect fluid with an
effective pressure given by the second of equations (\ref
{effective-matter-pressure}). Energy exchange is possible between the two
fluids. We have verified that equations (\ref{Friedmann3})--(\ref{Friedmann4}%
) reproduce successfully the known cosmological evolution in a variety of
models for the bulk geometry (Schwarzschild-AdS \cite{kraus}, Vaidya-AdS 
\cite{vaidya,tet2}, static perfect fluid in an AdS bulk \cite{tetap}, etc).

The equation of state ${p} _{\text{eff}}={p}_{\text{eff}}(\rho _{\text{eff}})
$ of the mirage component is not expected in general to have a simple form.
The reason is that $\bar{p} _{\parallel}$ is the pressure of the bulk fluid
perpendicularly to the brane, \emph{as measured by the brane observer}. The
only simple case involves an empty bulk energy-momentum tensor, for which 
$\bar{p} _{\parallel}=0$, ${p} _{\text{eff}}={\rho} _{\text{eff}}/3$. 
This is the well known case of mirage radiation with $\rho _{\text{eff}}=12M^3\mathcal{M}_0/(\pi ^2V \ell^4)$.

An interesting question concerns the possibility of having accelerated
expansion on the brane as a result of the brane-bulk interaction. The
Raychaudhuri equation (\ref{Raychaudhuri3}) provides important intuition on
this problem in a general framework. The acceleration parameter is
proportional to $\dot{H}+H^2$. For this to be positive one or more of the
following conditions must be satisfied: a) The effective cosmological
constant $\lambda$ is positive. b) The brane matter satisfies $\tilde{\rho}
< V$ and $\tilde{p}<-\tilde{\rho}/3$. c) The brane matter satisfies $\tilde{%
\rho} > V$ and $\tilde{p}<-2\tilde{\rho}/3$. d) The comoving mass $\mathcal{M%
}$ is negative. e) The pressure $\bar{p} _{\parallel}$ of the bulk fluid
perpendicularly to the brane, as measured by the brane observer, is
negative. In general negative pressures are associated with field
configurations. The possibility of a negative comoving mass seems
problematic at first sight, as usually it implies the existence of naked
singularities or instabilities. A counter example is a brane with negative
tension in the two-brane model of \cite{rs}.

\section{Acknowledgments}

One of the authors (NT) is partially supported through the RTN contract
MRTN-CT-2004-503369 of the European Union and the research program
``Kapodistrias'' of the University of Athens. Both authors acknowledge the
financial support of Ministry of National Education through the research
program ``Pythagoras'', grant no 70-03-7310.

\end{document}